\documentclass[showpacs,preprintnumbers,aps]{revtex4}
\usepackage{dcolumn}
\usepackage{bm}
\usepackage{epsf}
\usepackage{graphicx}

\input colordvi

\begin{document}

\title{Sensitivity of the neutron crystal
diffraction experiment to the neutron EDM \\and to the nuclear
P-,T-violating forces. }

\author{V.G. Baryshevsky and S.L. Cherkas}
\affiliation{Institute for Nuclear Problems, Bobruiskaya 11, Minsk
220030, Belarus}
\date{\today}

\begin{abstract}
We establish a link between an angle of the neutron polarization
rotation in a crystal diffraction experiment and constants of the
P-,T- violating interactions. The consideration applies to the
energy range of thermal and resonance neutrons.
\end{abstract}

\maketitle


Parity and time reversal violation in nuclear and atomic physics
 remains
the important contemporary problem. One of the way to search for
P-,T-odd interactions consists in measurement of the electric
dipole moment of an elementary particle. The most precision test
performed with ultra could neutrons in electric field gives
restriction $10^{-26}$ e cm for the electric dipole moment of a
neutron \cite{edm}. However others tests exist, namely, searching
for the neutron dipole moment by a crystal-diffraction method  was
suggested \cite{nat,for,for1,v,bar,bar1} and now is under
realization in the St. Petersburg Institute for Nuclear Physics
(PINP)\cite{fed,fed1}. The test is based on the neutron spin
rotation  in the non-centrosymmetruc crystal. Correction to the
neutron refractive index arises due to neutron rescattering by the
crystalline planes \cite{bar,bar1}.  Because  neutrons interact
with the nuclei of a crystal by the electromagnetic and strong
interactions a question arises about possible contribution of the
P-,T- violating nuclear forces to the  effect under consideration.
In this paper we analyze some different mechanisms by which the
P-,T-violating nuclear forces  can contribute to the neutron spin
rotation in a crystal.

According to  Refs. \cite{bar, bar1} angle of neutron polarization
rotation $\phi_{PT}$ in a crystal of a length $L$ is determined by
the spin-dependent part of the refractive index:
\begin{equation}
\phi_{PT}= 2 k L |\bm M_{PT}|,
\end{equation}
where $k$ is the neutron wave number,  $|\bm M_{PT}|$ is absolute
value of the vector forming the P-,T- violating matrix part of the
neutron refractive index $n_{PT}\sim \bm \sigma\cdot\bm M_{PT} $.
Vector  $\bm \sigma$  consists of Pauli matrices.

Under condition of a weak diffraction at some particular
crystalline planes corresponding to the vector $\bm \tau$ the
refractive index can be estimated as \cite{bar, bar1}
\begin{equation}
n_{PT}\sim \left(\frac{4\pi}{k^2 V}\right)^2\frac{f_s \,f_{PT}(\bm
\tau)}{2\alpha_B}\,e^{-2w(\bm \tau)}s(\bm \tau),
\end{equation}
where $f_s$ is the neutron scattering amplitude  due to strong
interaction, $f_{PT}(\bm q)$ is the P-,T-violating scattering
amplitude,
\begin{equation}
\alpha_{B}=\frac{ \bm \tau( \bm \tau+2\bm k)}{k^2}, \label{alphab}
\end{equation}
$V$ is a volume of a unit cell of the crystal,  $e^{-w(\bm
\tau)}=exp\left(-u^2 \tau^2/4\right)$ is the factor of
Debye-Waller and $u^2$ is a mean value of the square of the
amplitude of  the thermal motion  atoms about their equilibrium
positions. The multiplier $s(\bm \tau)=\sum_{jl}\left\{\exp({i\bm
\tau\bm R_{l}})-\exp({-i\bm \tau\bm R_{j}})\right\}$ \cite{com},
where summation is carried out over atoms in a unit crystal cell,
  characterizes
degree of violation of the central symmetry of the crystal cell
and $s(\bm \tau)$ is zero for crystal possessing center of
symmetry. That is, only non central symmetric (piezoelectric)
crystals to be used for the above diffraction experiments.

\begin{figure}[h]
\vspace{0. cm} \hspace{.5 cm}
 \includegraphics[width=4.5 cm]{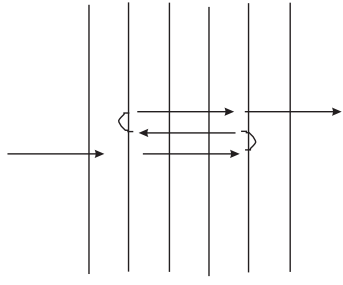}
 \vspace{. cm}
\caption{ Rescattering of a neutron by the crystalline planes.}
\label{ris}
\end{figure}

 In the geometry, where the wave vector of the  rescattered
wave is parallel to the incident one (see Fig. \ref{ris})
  no other P-,T- even
interactions are able to produce polarization rotation and the
presence of the last will be a signal of the P-,T- violation.

Let us compare contribution of the different sources of P-,T-
violation, namely, the P-,T- odd neutron nuclear interactions and
interaction of  neutron EDM with crystal electric field,  to the
angle of polarization rotation (or spin dichroism) which to be
measured in the current PINP experiment.

At first we consider  P-,T-odd neutron scattering amplitude by
nucleus.
 The amplitude contains
electromagnetic and nuclear parts. The first one is due to
interaction of the neutron dipole moment with the atom electric
field. The last one arises due to nuclear forces and can be
deduced from the P-,T- odd nucleon-nucleon interaction, which in
part can be described by the one pion exchange.

Restricting only to the $\pi$-mesons Lagrangian density of the
interaction can be written in the form \cite{finn}
\begin{equation}
{\mathcal L}=i g_\pi \bar N\gamma_5(\bm \tau\bm \pi)N+ \bar
g_\pi^{(0)}\bar N\,(\bm \tau \bm \pi)\,N +\bar g^{(1)}_{\pi}\bar
N\,\pi_0\,N+\bar g^{(2)}_\pi\bar N(3\tau^z \pi_0-\bm \tau\bm
\pi)N,
\end{equation}
where $N(x)$ is the nucleon field and $\pi_0(x)$, $\bm \pi(x)$ are
fields of $\pi$-mesons. Three last terms are parity and time
reversal violating.

\begin{figure}[h]
\vspace{0. cm} \hspace{.5 cm}
 \includegraphics[width=7.5 cm]{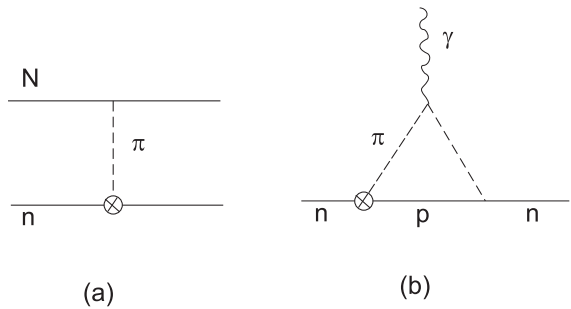}
 \vspace{. cm}
\caption{ (a) Diagram of one pion exchange, (b) Contribution of
the pion loop to the neutron EDM. P-,T- violation vertex is
denoted by a crossed circle.} \label{nucl}
\end{figure}

In the frame of the one pion exchange (see Fig. \ref{nucl} (a))
interaction of the incident neutron with the one of nuclei
nucleons
 can be deduced \cite{finn}:
\begin{eqnarray}
V_{PT}^{(\pi)}=\frac{1}{2 m_N m_\pi^2}\biggl(g_{\pi}\bar
g_{\pi}^{(0)}(\bm\tau\bm\tau_1)(\bm \sigma-
\bm\sigma_1)+g_{\pi}\bar g_{\pi}^{(1)}\left((\tau_1^z+ \tau^z)(\bm
\sigma-\bm \sigma_1)+(\tau_1^z- \tau^z)(\bm \sigma+\bm
\sigma_1)\right)\nonumber\\+g_{\pi}\bar
g_{\pi}^{(2)}(3\tau^z\tau_1^z-\bm\tau\bm\tau_1)(\bm\sigma-\bm\sigma_1)\biggr)\bm
\nabla\delta^{(3)}(\bm r),
\end{eqnarray}
where due to low energy of the neutron the interaction is
considered as having a zero range. Let us turn to the
neutron-nucleus interaction suggesting that the nucleus contain
equal number of non polarized proton and neutrons obeying the
distribution function $\rho(\bm r)$, which is normalized as $\int
\rho(\bm r)d^3\bm r=1 $.
 In this
suggestions the averaging gives the P-T-  odd neutron-nuclei
interaction:
\begin{equation}
V_{\mbox{\tiny NPT}}=-\frac{A}{ m_N m_{\pi}^2}g_{\pi}\bar
g_{\pi}^{(1)}\bm\sigma\bm \nabla\rho(\bm r), \label{single}
\end{equation}
where $A$ is the number of nucleons in nuclei, and relation $\int
\rho(\bm r^\prime) \nabla\delta^{(3)}(\bm r-\bm r^\prime)d^3 \bm
r^\prime=\nabla\rho(\bm r)$ is used.

 The
terms proportional to the $\bar g^{(0)}_\pi$ and $\bar
g^{(2)}_\pi$ vanishes because the isospin dependence leads to the
opposite sign for the $n-n$ and $n-p$ interaction whereas we
consider the nucleus  containing equal number of the $n$ and $p$.
In the the general case of the nucleus with the different numbers
of neutrons and protons contribution proportional to the constants
$\bar g^{(0)}_\pi$, $\bar g^{(2)}_\pi$ will be present.

 The next step is   the
evaluation of the neutron scattering amplitude by nucleus in the
first order on the momentum transferred $\bm q$
\begin{equation}
f_{\mbox{\tiny NPT}}(\bm q)=-\frac{m_N}{2 \pi}\int V_{\mbox{\tiny
NPT}}(\bm r)e^{-i\bm q\bm r}d^3 r=iA\frac{g_{\pi}\bar
g_{\pi}^{(1)}}{2\pi}\frac{\bm\sigma\bm q}{m_\pi^2}F(\bm q),
\label{npt}
\end{equation}
where one can set form factor $F(\bm q)$ to unity when the neutron
wave length much greater than the radius of a nucleus.

Another source of P- T- violation is the interaction of the
electric dipole moment of a neutron with the electric field of an
atom leading to the following interaction:
\begin{equation}
V_{\mbox{\tiny EDM}}=-d_n\,\bm \sigma\bm E=d_n\,\bm \sigma \bm
\nabla \left(\frac{Ze}{r}\exp(-r/R_a)\right)=-d_n Z\,e\,\frac{\bm
\sigma \bm r}{r}\left(\frac{1}{r^2}+\frac{1}{r
R_a}\right)\exp(-r/R_a),
\end{equation}
where $R_a$ is the atom radius.
 Scattering amplitude for this case
has the form
\begin{equation}
f_{\mbox{\tiny EDM}}(\bm q)=\frac{m_n}{2\pi} \,d_n\, Z\,e \int
\frac{(\bm \sigma \bm r)}{r}\left(\frac{1}{r^2}+\frac{1}{r
R_a}\right)\exp(-r/R_a)e^{-i\bm q\bm r}d^3\bm r= -2 \,i\, m_n\,d_n
Z e \frac{ (\bm \sigma\bm
q)}{q^2}\left(1-\frac{1}{1+q^2R_a^2}\right). \label{edm}
\end{equation}

It should be noted that the neutron dipole moment can be expressed
through $\bar g_{\pi}^{(0)}$ (see references in \cite{ann} and
Fig. \ref{nucl} (b))
\begin{equation}
d_n=\frac{e}{4 \pi^2 m_n}g_{\pi}\bar
g_{\pi}^{(0)}\ln\frac{m_n}{m_\pi}, \label{edm1}
\end{equation}
which allows to compare $f_{\mbox{\tiny EDM}}(\bm q)$  with
$f_{\mbox{\tiny NPT}}(\bm q)$. Finally we come to the conclusion
that the
 contribution of the neutron EDM interacting with the
electric field of the atom
  is much greater than that originated from the  P-,T- violating nuclear
  forces (compare fifth and seventh columns of the Table 1),
 when one assumes that $\bar
g_{\pi}^{(0)}$ and $\bar g_{\pi}^{(1)}$ are of the same order.

However we can consider another situation when the measurements
are carried out in the vicinity of compound nucleus p-resonance.
It is known that the P-odd and P- ,T- odd effects are greatly
enhanced in the vicinity of the p-resonance.

In the case of  p-resonance neutron scattering amplitude has the
form \cite{bun,flam}
\begin{equation}
f_{PT}^{res}(\bm q)\sim\frac{(\bm \sigma \bm q)
}{k^2}\frac{\gamma_p W_{ps}\gamma_s}{(E-E_p+i \Gamma_p)(E_s-E_p)},
\label{fre}
\end{equation}
where $\gamma_p$ is the capture amplitude of the p-resonance,
$\gamma_s$ is that of nearest s-resonance, $W_{sp}$ is the matrix
element of the P-,T- violating interaction between compound $s-$
and $p-$ states of the compound nuclei. Capture amplitude to the
$p-$resonance is suppressed compared to the $s-$resonance one:
$\gamma_p\sim (k R)\gamma_s$, where $R$ is the radius of the
nuclei. According to the  theory of the compound state reactions
the wave function of the compound state can be represented as a
sum of the shell-model functions: $\Psi_p=\sum_m^N C_{pm} \psi_m$,
where $\psi_m$ is the basic shell model functions and $N\sim
 10^6$ is the number of the principal components \cite{bun,flam}. The
coefficients $C_{pm}$ as well as $C_{sm}$ are random giving the
following estimate for the matrix element between compound states
$W_{sp}\sim W/\sqrt{N}$ where $W$ is the typical value of the
single particle P-T- odd interaction, which according to the
equation (\ref{single}) can be estimated as $W\sim
\frac{g_{\pi}\bar g_{\pi}^{(1)}A}{m_n m_\pi^2 R^4} $. Note, that
not only the single particle interaction (\ref{single}) give rise
to the contribution to the matrix element, but also original
two-particle interaction. Thus, the terms proportional $\bar
g_{\pi}^{(0)}$ and $\bar g_{\pi}^{(2)}$ can also give contribution
of the same order to the matrix element $W$. Taking into account
that the typical energy interval between compound states
$E_s-E_p\sim \Delta E / N$, where $\Delta E\sim 7~MeV$ is the
typical energy distance between shell model states
 we come to the
 estimate
\begin{equation} f_{PT}^{res}(\bm q)\sim\frac{(\bm \sigma \bm q)
}{k^2}\frac{1}{k R}\frac{\sqrt{N}\,W}{\Delta
E}\frac{\gamma_p^2}{\Gamma_p} \label{esone}
\end{equation}
in the vicinity of p-resonance $E-E_p\sim \Gamma_p$.

Another way to estimate P-,T- odd resonance amplitude is to use
results of the measurement of P-violating, T-conserving spin
rotation in $^{139}La$ \cite{alf,yp,ser,vesna}. In these
experiments it was found
 the relation of the weak matrix element $\widetilde{W}_{sp}$ between compound
states to the energy difference of the p-resonance with the
nearest s-resonance
\begin{equation}
\frac{\widetilde{W}_{sp}}{E_s-E_p}=\frac{1.7\times
10^{-3}}{38}=4.5\times 10^{-5}.
\end{equation}
In assumption that this matrix element originates from the $\pi$
-meson Lagrangian including strong and P- odd, T-even interaction
\cite{mus}
\begin{equation}
\mathcal L=i g_\pi \bar N\gamma_5(\bm \tau\bm \pi)N+h_{\pi}\,\bar
N (\bm \tau\times\bm \pi)_3 N,
\end{equation}
leading to the P-odd, T-even two particle interaction
\begin{equation}
V_{\mbox{\tiny P}}(\bm r)=i\frac{g_\pi h_\pi}{2\sqrt{2}m_N
m_\pi^2}(\bm \tau_1\times \bm \tau_2)_3(\bm \sigma_1+\bm
\sigma_2)[(\bm p_1-\bm p_2),\delta^{(3)}(\bm r)],
\end{equation}

one may suggests the same coefficient of proportionality between
the coupling $g_\pi h_\pi$ and matrix element
$\widetilde{W}_{sp}$, and between the P-,T- violating  coupling
$g_\pi \bar g_\pi^{(i)}$ and matrix element $W_{sp}\,$. This
allows  to estimate $W_{sp}$ directly:
\begin{equation}
W_{sp}=\frac{ \bar g_\pi^{(i)}}{ h_\pi}\widetilde{W}_{sp},
\label{dir}
\end{equation}
where  $h_\pi$ can be estimated as $h_\pi=1.9\times 10^{-7}$
\cite{mus} (it can be compared with the value $g_\pi\approx 13$).
Other quantities are total neutron width of p-resonance
$\Gamma_p=0.045~eV$, neutron width $\Gamma_p^n=3.6\times
10^{-8}~eV$ \cite{rev} and that for s-resonance $\Gamma_s^n=0.1
~eV$. Energy distance to the s-resonance state is $E_s-E_p\sim
38~eV$ \cite{alf}.

 Let us come to the estimates of the polarization rotation
angle
 in the
different range of energies. For the parameter $\alpha_B$
(\ref{alphab}) describing deflection from the exact diffraction
condition in the backscattering geometry (Bragg angle equals
$180^o$) we have $\tau=2 k$ and
\begin{equation}
\alpha_B=4\left({\Delta k}/{k}+\Delta\theta^2\right).
\end{equation}
 The quantity $\Delta \theta \sim 10^{-3}$
describes typical angle spread around the Bragg direction and
$\Delta k/k\sim 10^{-6}$ describes wave number spread. Other
parameters are $L=0.5~ m$, $u=0.1~ \AA$, $V^{1/3}=5~ \AA$,
$R_a=2~\AA$, $R=1.45 A^{1/3}~fm$, where $A=139$ is the mass number
of nuclei and $Z=57$ is charge number. The value of the factor
$s(\bm \tau)$ describing the absence of the center of symmetry was
set to unity in the above calculations because the in the most of
the piezoelectric crystals central symmetry is violated strongly.
Value of the neutron EDM is taken $10^{-26}~e~cm$ and
corresponding product of the constants $g_{\pi}\bar g_{\pi}^{(i)}$
is determined from the equation (\ref{edm1}) in suggestion that
all the P-,T- violating constants $\bar g_{\pi}^{(1)}$, $\bar
g_{\pi}^{(2)}$, $\bar g_{\pi}^{(3)}$ are of the same order.

\widetext
\begin{table}
\caption{Angle of neutron spin  rotation due to interaction of the
neutron EDM with the crystal electric field --- $\phi_{EDM}$, due
P-,T- violating nuclear forces under potential scattering
--- $\phi_{NPT}$ and under resonance scattering --- $\phi_{res}$.
 The quantities marked by
the letter $^{a)}$ corresponds to the estimate (\ref{esone}),
whereas the case marked with the letter $^{b)}$ corresponds to the
Eqs. (\ref{fre}), (\ref{dir}). }
\label{tab1}       
\begin{tabular}{llllllll}
\hline\noalign{\smallskip}
E, eV &$\Delta \theta$&$\Delta k/k$&$\frac{4\pi}{k^2 V}f_s$& $\phi_{EDM}$ & $\phi_{res}$& $\phi_{NPT}$  \\
\noalign{\smallskip}\hline\noalign{\smallskip}
0.003 & $10^{-3}$&$10^{-6}$&$5.6\times 10^{-6}$&$4.7\times 10^{-5}$  & --- &$2.8\times 10^{-11}$  \\
0.1 &$10^{-3}$&$10^{-6}$&$1.7\times 10^{-7}$&$ 3.5\times 10^{-8} $& ---&$6.7\times 10^{-13}$\\
0.73 &$10^{-3}$&$10^{-6}$&$2.4\times 10^{-8}$& $1.4\times
10^{-10}$ &$ ^{a)}\,8.3\times 10^{-9}$&$2.0\times 10^{-14}$ \\
&&&&&$ ^{b)}\,5.6\times 10^{-10}$&\\
0.73&$10^{-4}$&$10^{-8}$&$2.4\times 10^{-8}$& $1.4\times 10^{-8}$ &$^{a)}\,8.3\times 10^{-7}$&$2.0\times 10^{-12}$\\
&&&&&$ ^{b)}\,5.6\times 10^{-8}$&\\
 \noalign{\smallskip}\hline
\end{tabular}
\end{table}
\widetext

 It should be noted that from the one hand the angle $\Delta \theta$ should be
greater then the mosaism of a crystal (typical value
$10^{-3}-10^{-4}$) and from the other hand it should satisfy
 the condition of a weak
 diffraction: $\alpha_B=4(\Delta \theta^2+\Delta k/k)>>\frac{4 \pi}{k^2 V}f_s$.

To summarize we have compared direct neutron EDM contribution and
that of the P-,T-violating nuclear forces to the angle of neutron
polarization rotation in the crystal diffraction experiment. It
turns out to be that at energies near $0.003$ eV only the first
one is considerable, (see first line of the Table 1)  and the
total number of neutrons needed is $N_{tot}=1/\phi_{EDM}^2\sim
4.5\times10^8$. Corresponding accumulation time is $T=N_{tot}/(N_0
S \frac{\Delta \theta}{\pi} 2\frac{\Delta k}{k})\sim 0.8 ~sec$
under the neutron flux $N_0=10^{15}~~neutrons/cm^2/sec$ at the
reactor zone (i.e. at the bottom of the reactor channel), area of
the crystal $S=30\times 30~cm^2$. It should be noted that we have
considered the best conditions, that is, the greatest possible
neutron flux and the large area crystal containing relatively
heavy La atoms (scattering length $f_s=8.2$ fm), whereas at
present time only large perfect crystals of quartz containing
light elements of periodic table  are available.

At higher energies, under resonance scattering, contribution of
the P-,T-violating nuclear forces begin to dominate,  however,
 at the energy 0.73 eV of the $^{139}La$ resonance one needs
$N_{tot}=1.5\times 10^{12}$ (see last line of table 1, case
$^{a)}$) and accumulation time $T=30$ days under the same flux and
crystal area. Hence the experiment in the resonance range should
be done at the reactors having much excess of the neutrons with
energies near $0.73$ eV.

The work is supported by the Belarus fund for fundamental
research, grant $\Phi06P-074$.

\end{document}